\documentclass[preprint]{aastex} 

\usepackage[dvips, bookmarks, colorlinks=true, pdftitle={Terrestrial
    Planet Formation from an Annulus}, pdf author={Kevin Walsh},
  pdfsubject={Terrestrial planet formation from an annulus},
  pdfkeywords={Planet formation, Annulus}]{hyperref}
\usepackage{graphicx} 
\usepackage{amsfonts} 
\newcommand{\hide}[1]{} %

\usepackage{wasysym}

\usepackage{hyperref}

\begin{document}
\title{Planetesimals to Terrestrial Planets: collisional evolution amidst a dissipating gas disk }
\author{Kevin J. Walsh \&  Harold F. Levison}
\affil{Southwest Research Institute, 1050 Walnut St. Suite 300, Boulder, CO 80302, USA}
\email{kwalsh@boulder.swri.edu}

\begin{abstract}

We present numerical simulations of terrestrial planet formation
that examine the growth continuously from planetesimals to planets in
the inner Solar System.  Previous studies show that the growth will be
inside-out, but it is still common practice to assume that the entire inner disk
will eventually reach a bi-modal distribution of embryos and
planetesimals. For the combinations of disk mass, initial planetesimal
radius and gas disk lifetime explored in this work the entire disk
never reaches a simple bi-modal mass distribution.

We find that the inside-out growth is amplified by the combined
effects of collisional evolution of solid bodies and interactions with
a dissipating gas disk. This leads to oligarchic growth never being
achieved in different places of the disk at the same time, where in
some cases the disk can simultaneoulsy support chaotic growth and
giant impacts inside 1~au and runaway growth beyond 2~au. The
planetesimal population is efficiently depleted in the inner disk
where embryo growth primarily advances in the presence of a
significant gas disk. Further out in the disk growth is slower
relative to the gas disk dissipation, resulting in more excited
planetesimals at the same stage of growth and less efficient
accretion. This same effect drives mass loss due to collisional
grinding strongly altering the surface density of the accreted planets
relative to the initial mass distribution. This effect decreases the
Mars-to-Earth mass ratios compared to previous works with no
collisional grinding. Similar to some previous findings utilizing
vastly different growth scenarios these simulations produce a first
generation of planetary embryos that are stable for 10-20~Myr, or 5-10
e-folding times of the gas dissipation timescale, before having an
instability and entering the chaotic growth stage.

\end{abstract}

\section{Introduction}

A long-standing issue in models of terrestrial planet formation is the
order of magnitude difference in mass between Earth and Mars.  Most
models aiming to track the final stages of planet formation start from
an intermediate stage of growth, with similar-sized planetary embryos
amidst a sea of planetesimals. These models regularly fail to recover
the relatively small mass of Mars for the simple reason of the surplus
of material at 1.5~au in most nominal surface density profiles -
leading to Mars-analogs that are usually 5-10 times too massive
\citep{Chambers:2001p7618,Raymond:2009p11530,Fischer:2014p13097}.
Some works have been able to create small Mars-analogs, but require
migrating planets or disk incontinuities to vastly deplete certain
regions of the disk
\citep{Hansen:2009p8802,Walsh:2011p12463,Izidoro:2014p15200,Clement:2018}. Other
models rely on the majority of the initial solid mass to remain in
cm-sized ``pebbles'', which can lead to a vastly different accretion
mode \citep{Levison:2015p20212,Morbidelli:2015p23049}. However, a
statistical simulation to this problem suggests that mass loss due to
collisional fragmentation and drift can make planetary embryos that
are a good match for Mars' mass and accretion timescale, but only for
relatively high disk masses and smaller initial planetesimals
\citep{Kobayashi:2013p11732}.

Complete models of terrestrial planet growth from planetesimals are
not yet fully explored, as the complicated interaction of collisional
fragmentation and gas disk affects have not been exhaustively explored
in numerical models spanning the entire growth from planetesimals to
planets.  Modeling the entirety of planet formation represents a
dynamic range problem where a huge population of dust becomes very few
planets, interactions between dust happens on very short timescales,
and the final accretion of the planets takes 10's of millions of
years.  These are the driving reasons that push the vast majority of
studies to take a piece-wise or regional approach to the problem, and
these efforts have produced a general picture of distinct stages of
growth.

The first stage of growth following planetesimal formation is
``runaway growth''. Owing to the effects of gravitational focusing,
the largest planetesimals in a given local region grow at a faster
rate than smaller neighbors, allowing them to runaway in mass relative
to the others \citep{Greenberg:1978p15127,Lissauer:1987p21583}. This
leads to a break in the size distribution that is secured once the
largest bodies begins to stir the orbits of their smaller neighbors,
increasing relative velocities and discouraging further growth among
the population of small bodies. This transition into ``oligarchic
growth'' occurs when the large bodies attain roughly half the total
local mass
\citep{Kokubo:1998p9706,Kokubo:2000p10143,Chambers:2006p11697}.

With growth slowed and mass partitioned evenly between $\sim$10--100
large bodies and numerous small bodies, the oligarchic growth stage
serves as the typical starting point for numerical models of the full
terrestrial planet region. This will continue until a final set of
planets have been built by way of giant impacts during the ``giant
impact'' or ``chaotic growth'' stage of growth where the embryos'
orbits eventually become crossing and giant embryo-embryo collisions
occur due to the lack of damping from planetesimals or the gas disk
\citep[see][]{Morbidelli:2012p11505}. The expectation is that the
spacing of the oligarchs, or embryos, is constant in terms of their
mutual Hill Spheres but increasing with mass with further distance for
typical surface density profiles. No simple prescription exists for
the location and mass of the planetesimal population and it is
typically assumed to follow the initial surface density profile
\citep{Obrien:2006p8571,Raymond:2009p11530}.


While each of the stages of growth have been explored in detail,
combining them into one model of planet formation relies on
extrapolations from very regional simulations to establish conditions
for global simulations. A key aspect to this extrapolation is to
assume that oligarchic growth can be reached throughout all regions of
the disk at the same time, which allows modelers to start modeling the
final stage of growth from a disk of embryos amidst
planetesimals. Since observed gas disk lifetimes
\citep{Haisch:2001p10160} are similar to estimates for planetary
embryo growth timescales at 1~au (2-10~Myr) it is common to ignore
many of the effects of the gaseous solar nebula, which can strongly
affect the orbits of both the small and large bodies at oligarchic
growth stage. Finally, collisional fragmentation is not typically
modeled throughout as it is numerically challenging due to the
possible rapid increase in the number of simulation particles if/when
objects start fragmenting. This is an important omission as fragments
can dynamically interact with large ones, changing orbits and
accretion efficiency locally, or simply be lost due to rapid gas drag
or at later times by Poynting Robertson drag
\citep{Kenyon:2006p11683,Leinhardt:2009p10318,Chambers:2013p19990,Carter:2015p19516}.

The aim for this work is to model the growth from planetesimals in the
inner Solar System and to present simulation results that continue
through the entire process to systems of planets. The modeling
techniques will include the continuous growth from planetesimals to
planets in the midst of the dissipation of the gaseous solar nebula
and collisional evolution. The model extends through the stage of
chaotic growth and giant impacts, and multiple simulations are
performed for a range of initial conditions to provide a statistical
comparison of the produced planetary systems.

\subsection{Previous studies - planetesimals to planets}

Previous studies have either focused on the early growth from
planetesimals to embryos or on the latter stages from embryos
to planets. One of the foundations of most analytical approaches is
establishing the end point of the early stages of growth, before
chaotic interactions between embryos should begin, as a function of
disk properties. This end, and the produced suite of embryos or
``oligarchs'', typically sets up the initial conditions for numerical
models that are capable of modeling the more chaotic final stages of
growth. Aspects of these works provide a solid foundation that should
be re-created in any end-to-end modeling effort.

A near universally used concept is that the largest body at any
distance should grow to a so-called Isolation Mass ($M_\mathrm{iso}$),
which is defined as the total mass per well-separated feeding zone
\citep{Lissauer:1987p21583,Kokubo:1998p9706,Kokubo:2000p10143,Kobayashi:2013p11732}.
This depends on the spacing of planetary embryos, but with the
typically assumed $\sim$10 Hill sphere spacing
\citep{Kokubo:1998p9706} the isolation mass as a function of semimajor
axis and mass of the disk can be easily determined:

$$M_{\mathrm{iso}}=0.13\chi^{3/2}\Big(\frac{a}{1.5\mathrm{au}}\Big)^{3}M_{\oplus}$$,

\noindent where $\chi$ is the scaling of the classical Hayashi Minimum
Mass Solar Nebula (7.1~g~cm$^{-2}$ surface density for solids and
1.7$\times$10$^{3}$~g~cm$^{-2}$ for gas at 1au) and $a$ is the
semimajor axis of the growing embryo \citep{Hayashi:1981p22959}. This
accounts only for accretion of local planetesimals by embryos, and
does not consider mass re-distribution (drifting or scattered
planetesimals), embryo-embryo accretion or the final giant impact
stage of planet growth (note that \citealt{Chambers:2006p11697}
include an increase of 50\% to $M_{\mathrm{iso}}$ to account for
embryo-embryo mergers in a semi-analytical model).

\begin{figure}[!h]
\includegraphics[angle=-90,width=3in]{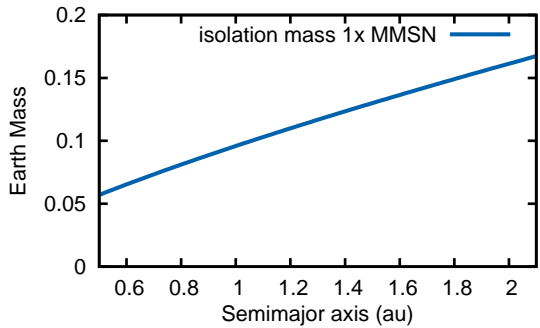}
\caption{The theoretical isolation mass for a Minimum Mass Solar Nebula assuming embryos spaced by 10 Hill Spheres \citep{Kobayashi:2013p11732}.}
\label{fig:Iso}
\end{figure}

For a typical surface density that scales as $a^{-3/2}$ the isolation
mass increases with distance, where at 1.5~au this leads to typical
embryos roughly equal to a Mars mass (see Figure \ref{fig:Iso}).  This
concept encodes no information about timescales and does not consider
mass loss and re-distribution due to collisional fragmentation. Mass
loss due to fragmentation and drift will depend strongly on the
behavior of the gas disk over time
\citep{Kominami:2002p21136,Chambers:2008p78,Kobayashi:2010p9661,Ormel:2010p11711}. Despite
these clear inadequacies, it does serve as a foundation for many of
the analytic approaches below and as a valuable fiducial for comparing
the early stages of growth found in numerical approaches.

\subsubsection{Analytical Descriptions including Fragmentation and Gas Effects}

Semi-analytic models of embryo growth necessarily consider the
collision probability between the growing bodies and the local swarm
of smaller planetesimals, where the dynamical excitement of all bodies
and their size distribution is a critical component in determining
accretion rates.  The fragmentation and drift of planetesimals
complicate this calculation, and where included has been found to
alter the accretion rates due to the loss, or re-distribution, of mass
\citep{Chambers:2006p11697,Kobayashi:2010p9661,2012ApJ...747..115O}. 

\citet{Chambers:2008p78} included this effect in a semi-analytic
approach to modeling the growth of planetary embryos but was
restricted in the handling of an evolving size distribution of
planetesimals, where fragmented planetesimals all were put in one size
bin. This work also considers embryo-embryo mergers and embryo
atmospheres, for which the former affect is found to roughly increase
embryo growth rates by the simple flat rate of $\sim50$\%.  With all
of the above physics included, and assuming a decreasing surface
density of the gas disk with an exponential 2~Myr timescale, and
planetesimals initially with diameters of 10~km
\citet{Chambers:2006p11697} finds inside-out growth of embryos, with a
slow march towards isolation masses over 3-10~Myr (see
\citealt{Chambers:2006p11697} Figure 12). For nominal MMSN masses of
gas and solids and surface density profiles, the embryo growth is
found to approach isolation masses at 1~au in under 1~Myr and 2~au in
$\sim$3~Myr and have a growth timescale that scales as $t^3$, with the
only fluctuations/variations related to initial planetesimal
excitement briefly driving oligarch growth away from that
scaling. Other noted affects are that shallow surface densities lead
to growth exceeding isolation mass because more solid material drifts
into regions of the inner disk (from outside the inner disk) than
drifts out of it. Increased masses increase the isolation mass and
decrease the time to reach isolation mass.

It has been found that growth timescales also depend on the
initial size of the planetesimals
\citep{Kobayashi:2010p9661,Kobayashi:2013p11732}, with more rapid
growth and more collisional grinding for smaller initial sizes. When
planetesimals can collisionally evolve and create a collisional
cascade down to small sizes, the timescale for depletion by
fragmentation becomes inversely proportional to the surface density
and increases for increasing initial planetesimal size and scales
strongly with semimajor axis ($\sim a^{3.2}$) (see
\citealt{Kobayashi:2013p11732} Eq. 4.). Therefore embryo masses will be
below isolation masses when the initial planetesimal size is less than
$R_\mathrm{init}\lesssim100$~km. When initial planetesimal sizes are
very small, 1--10~km, embryo growth by fragment accretion becomes more
important, where, with no radial drift of fragments, embryos can reach
isolation mass more rapidly than growth from the initial planetesimals
population.

\citet{Kobayashi:2013p11732} propose that a solution to the combined
problem of the mass of Mars and its rapid formation timescale can be
addressed by decreasing initial planetesimal size to increase
accretion rate, and to also increase the surface density to make up
for mass loss to collisional grinding. Combined initial radii between
1--10~km and surface densities between 2--3 MMSN may satisfy both
criteria -- and this is tested numerically in this work.

Despite significant advantages to modeling growth analytically during
runaway and oligarch growth stages there are strict limitations once
oligarchs start crossing orbits, having chaotic orbital evolution or
altering the mass distribution of material by scattering
planetesimals. Hence we cannot draw from these works conclusions about
the final planetary system, in terms of the number of planets, orbital
properties or timescales for the giant impacts that serve as critical
chronometers for growth and evolution timescales.

\subsubsection{Numerical Descriptions including Fragmentation and Gas Effects}

Numerical efforts typically rely on statistical techniques to account
for the short timescales and large numbers of interactions between
very small objects in a gas-disk and then necessarily transition to
$N$-body integrations of planetary embryos having distant
gravitational encounters for millions of years. Combining these very
different types of calculations results in hydrid $N$-body codes with
a range of capabilities
\citep{Kenyon:2006p11683,Leinhardt:2009p10318,Levison:2012p12338,Morishima:2015p20942}

Numerical models typically find the inside-out growth predicted by the
analytics. In hybrid $N$-body simulations with
$R_\mathrm{init}=1-3$~km between 0.86-1.14au and a $\sim$1$\times$MMSN
surface density, \citet{Kenyon:2006p11683} find that oligarchs
($m>10^{26}$~g) are produced at the inner edge of the annulus in
300,000 years with the transition to oligarchy moving outwards across
the annulus. The timescales for reaching and leaving oligarchic
growth is found to depend strongly on initial surface density, where
chaotic growth regime begins earlier and is more violent for more
massive disks. When modeling a disk stretching from 0.2-2~au they find
inside-out growth with oligarchs formed on the inner edge in
$\sim$100,000 years and in 1~Myr at the outer edge. No fragmentation
or mass loss due to collisional grinding was included, and the
dependence on gas disk timescale for onset of the chaotic stage of
growth was not investigated.

The dependence on the timescale and nature of the dissipating gas-disk
was the focus of a study utilizing a full $N$-body model of the final
stages of planet formation \citep{2004Icar..167..231K}. The onset of
embryo crossing orbits depends strongly on the gas disk lifetime and
allows for continued accretion of embryos before giant impacts
initiate (see also \citet{Walsh:2016p23062} which found similar
outcomes while modeling a different formation scenario). The
implication is that the gas density is more closely correlated with
the onset of embryo-crossing orbits than are the damping affects of
dynamical friction from planetesimals. Therefore attempting to model
just the final stages of growth it may be admissable to begin a model
under the assumption that the gas has just dissipated, allowing
crossing embryos orbits, but there would be no clear guidance on what
to assume in the planetesimal population. A long gas lifetime could
keep a stable suite of embryos around for a long time, but would allow
for more depletion of planetesimals relative to a short disk
lifetime.

Collisional fragmentation and mass loss has been modeled in full
$N$-body simulations utilizing a specific state-of-the-art collision
model, called EDACM \citep{Leinhardt:2012,Leinhardt:2015p18851},
integrated into the $N$-body accretion model $pkdgrav$
\citep{Carter:2015p19516,Leinhardt:2015p18851}. This model utilizes
particle radius inflation to decrease computational requirements and
is particularly well-suited for very high-resolution simulations
starting mid-way through runaway growth and continuing into the onset
of giant impacts and chaotic growth. They are not ideal for the final long
runout of accretion due to the second-order integrator accuracy.
These works started with similar sized particles
\citep{Leinhardt:2015p18851} or from with a size distribution of
planetesimals representative of a disk still in runaway growth ranging
up to 0.01 Earth Masses for the largest
\citep{Carter:2015p19516}. Both note the inside-out growth
characterized by an ``embryo front'' that moves through the disk.  The
growth to embryos in the midst of the gas disk produced less massive
embryos, but on shorter timescales, than for systems with no gas drag
(see also \citealt{Wetherill:1993p11515,Kokubo:2000p10143}).

When bouncing or fragmention is included for the final stages of
growth, during the epoch of giant impacts and the final growth of the
planets, the primary affect is found to extend the timescale for
formation and produce slightly less dynamically excited final systems
of planets \citep{Kokubo:2010p20928,Chambers:2013p19990}. These works
either allow bouncing of embryo's to simulate non-perfect accretion
scenarios \citep{Kokubo:2010p20928} or set a minimum fragment size
that can be produced in an energetic collision
\citep{Chambers:2013p19990}. Modeling this effect is particularly hard
as large impacts could create a huge number of new particles spanning
all sizes all the way down to dust, instantly increasing the $N$ of a
simulation.

\subsubsection{This work}

The combined previous research finds numerous important affects that
should be expected to strongly impact the growth and final suite of
planets. Many of these are not typically considered when constructing
initial conditions for models of the final stages of planet growth:

\begin{enumerate}
\item Growth timescales should be increase with distance,
\item Fragmentation can increase the accretion timescale and when
  paired with small initial planetesimals will decrease embryo masses,
\item The state and dissipation of the gas disk may determine the
  onset of giant impacts and chaotic growth,
\end{enumerate}

In this work we aim to explore the growth of the terrestrial planets
from planetesimals to planets, in the midst of a decaying gas disk, as
a function of the gas disk decay timescale and with a size
distribution of planetesimals that collisionally evolve and can be
lost due to collisional fragmentation. We endeavor to compare, where
possible, with the timescales and scaling relationships previously
discussed and with results from different analytical and numerical
approaches. Specific questions to be answered:

\begin{enumerate}
\item When and where is a disk in oligrachic growth?
\item What is the spatial and size distribution of planetesimals
  during this growth?
\item How does exponential disk decay affect embryos and planetesimals
  at different stages of growth?
\end{enumerate}

Another driving goal of this work is to develop a prescription to
describe the state of the inner disk growth as a function of time, gas
disk properties and initial planetesimal properties. The constant
mention that previous initial conditions are typically inadequate is
meant to motivate future modelers to use something superior, and we
aim to facilitate this.

Futhermore, by the nature of the modeling techniques used, we will
present final planetary systems for numerous sets of input
parameters. In places we present suites of simulations to give a more
robust view of possible outcomes. Naturally, this covers a small
region of large parameter space, but provides the important connection
between models of embryo growth in the context of important
constraints on final planetary systems.

\section{Methods}

To do planetesimal to planet simulations we use the code {\tt LIPAD},
which stands for Lagrangian Integrator for Planetary Accretion and
Dynamics \citep{Levison:2012p12338}. {\tt LIPAD} is based on the
algorithm known as the Wisdom-Holman Mapping (WHM;
\citealt{Wisdom:1991p456}). It can treat close encounters between
bodies using the algorithms of {\tt SyMBA}
\citep{Duncan:1998p7713}. SyMBA also allowed for increased
  computational speed by breaking the population into larger
  ``embryos'' that had full $N$-body gravitational interactions with
  each other, and a smaller population of planetesimals that had
  $N$-body gravitational interactions with embryos, but not each
  other. Here, {\tt LIPAD} utilizes ``tracer'' particles that each have the
identical total mass for calculating the gravitational evolution of
the system.  Each tracer also has a representative radius $s$, such
that it represents a swarm of particles each of radius $s$ whose mass
sums up to the total mass of the tracer itself.  The tracers have
  normal $N$-body interactions with larger embryos (as in SyMBA), but can
  evolve the sizes and random velocities of other tracers.

Collisional probabilities are calculated for each tracer as a function
of its size $s$ and the total mass, sizes and orbits of its
neighboring tracers. The dynamics of each tracer is modeled with
direct gravity calculations with all embryos as well as other
dynamical effects (dynamical friction, viscous stirring between
tracers), where many calculations depend on the particle's radius $s$
and the masses, sizes and orbits of its neighbors. Tracer
  particles can grow to the point where they are promoted into being
  embryos, although they go through a phase as ``sub-embryo'' so as to
  avoid unphysically large scatterings as they would be only slightly
  more massive than the tracer particles
  (see\citep{Levison:2011p17955} for more details).

There are limits on the smallest size that tracers can reach. As they get smaller, collisional probabilities increase and simulations become more computationally expensive. Thus a small size is set, below which particles are removed from the system, under the presumption that they will collisionally grind to dust on rapid timescales or, if the gas disk is still present, they will experience rapid inward drift and remove themselves from affecting the larger dynamics of the system. In the presence of the gaseous nebula the sizes of particles that experience maximum inward drift is near 1~meter, which is selected as the smallest size allowed in the collisional cascade for all simulations presented in this work. Tests were run with a small size of 0.1~m, for the first 1~Myr of evolution and then from 4--8~Myr of evolution of a nominal simulation, and while the mass loss amounts and locations were not identical, they were similar enough to suggest that when simulations with smaller sizes are computationally feasible the results will look quite similar.

For the first 4~Myr
of each simulation the giant planets were at 3.5 and 6~au with 1 Earth
mass each. At 4~Myr they were moved instantaneously to 5.0 and 9.2~au
and increased to their current masses on dynamically cold orbits
($e=i=0$), where these orbits are meant to represent the
low-eccentricity orbits preceeding the expected giant planet
instability \citep{Levison:2011p17955}.

The effects of the solar nebula are modeled through aerodynamic drag
and type-I eccentricity damping (see \citealt{Levison:2012p12338} and
references therein). For Nominal tests here the gas disk started with
a surface density at 1~au of 1.4$\times10^{-9}$~g~cm$^{-3}$
\citep{Hayashi:1981p22959}, and the density profile with radius and
time is $\rho(r,t) = \rho_0r^{-\gamma}\exp(-t/\tau)$, where $\gamma$
is 9/4 and the decay timescale $\tau$ is 2~Myr. The disk also has a
flaring profile $z = h\Big(\frac{r}{\mathrm{AU}}\Big)^f$ where the scale height
$h$=0.05~au, and a power law profile $f$=1.25. The mass of the gas
disk, the population of solids and the timescale for disk decay were
all varied in different suites of simulations.

Planetesimals and embryos had a density of 3~g~cm$^{-3}$ throughout.
When a collision occurs a fragmentation law is used to determine the
outcome \citep{Benz:1999p505}. This determines the expected size
distribution of fragments, and a radius $s$ is chosen from that
distribution for each tracer involved. The size distribution of the
system requires the inclusion of many tracer particles, each with
different sizes to build a size distribution regionally (see
\citealt{Levison:2012p12338} for more detail).  The handful of
  very large, embryo-embryo, impacts in each simulation are treated in
  the exact same fashion, despite \citet{Benz:1999p505} experiments
  being designed for impacts between bodies with radius $<$ 1000~km
  (see Deienno et al. in prep for further work on energy dissipation
  in giant impacts in LIPAD simulations).

 Simulations utilized a range of resolution, total number of initial
 tracer particles, and semimajor axis range, but always had an inside
 edge of the mass distribution at 0.7~au.


\subsection{Nominal Case}

The ``Nominal'' case uses 5000 particles starting with
$R_\mathrm{init}=30$~km with a 1-$\sigma$ spread in radius of 3~km.  The initial
surface density profile follows $r^{(-3/2)}$ for a total mass of solid
material equaling 3.32 Earth masses between 0.7-3.0~au.  A gas disk is
included in the calculation that has an initial gas density of
1.4$\times10^{-9}$ g/cm$^{3}$ at 1~au that provides aerodynamic gas
drag and type-I eccentricity damping on particles
\citep{1976PThPh..56.1756A,Tanaka:2004p8662}. The gas density
decreases exponentially with a 2~Myr timescale. The smallest allowable
particle produced in a collision is 1 meter, below that size particles
are removed from the simulation, and tracked as mass lost to
collisional grinding.
  
The Nominal and the related high-resolution case (with 2$\times$ the
number of tracer particles, but identical total system mass), were
both used for analysis of initial growth to $\sim$10~Myr, and the
related timescales associated with the transition to oligarchic growth
as a function of semimajor axis. These cases were not used in the
analysis of the final planetary systems due to prohibitively long
run-times - all final planetary systems were part of the various
low-resolution suites of simulations that used 2000 tracer particles
between 0.7--2.25~au.

\subsection{Grid of Mass and Gas Lifetimes}
We rely below on a grid of simulations where increasing total disk
masses are paired with decreasing disk lifetimes. Disk masses of
2$\times$, 1.5$\times$ and 1$\times$ Minimuim Mass Solar Nebula 
were modeled with gas disk dissipation timescales of 1~Myr, 2~Myr and
3~Myr respectively for a total of three different sets of
parameters. For these three cases, four simulations each were run
  with randomized orbital elements and sizes, and are combined for
some analyses. Additionally, there are other singular test cases, such
as simulations with no gas disk, gas disks that extend forever, no
collisional grinding etc., that are used as simple tests of different
scalings. For all of these cases, where the simulations were run to
115~Myr, they were low-resolution, with 2000 initial tracers between
0.7--2.25~au.

\subsection{Test of Kobayashi \& Dauphas 2013}

A direct test of a proposed set of disk properties that could be
capable of growing Mars-sized embryos fast enough to match Mars'
growth timescale was put forth in a series of analytical arguments and
statistical simulations \citep{Kobayashi:2013p11732}. Here,
four simulations were run using $R_\mathrm{init}=5$~km with a
2$\times$MMSN disk mass and a 2~Myr disk dissipation timescale, where
the parameters were selected from Figure 3 of
\citealt{Kobayashi:2013p11732} as being safely in the region
satisfying expected embryo sizes and growth timescales. These
simulations presented here otherwise utilized the Nominal
settings described above, with the exception of not including massive
giant planets during any of the growth (instead these simulations
include 0.1 Earth Mass planets at 5.4 and 9.0~au - a difference we do
not expect to be important for the tests reported here).

\section{Results}

\begin{figure}[!h]
\includegraphics[angle=-90,width=6in]{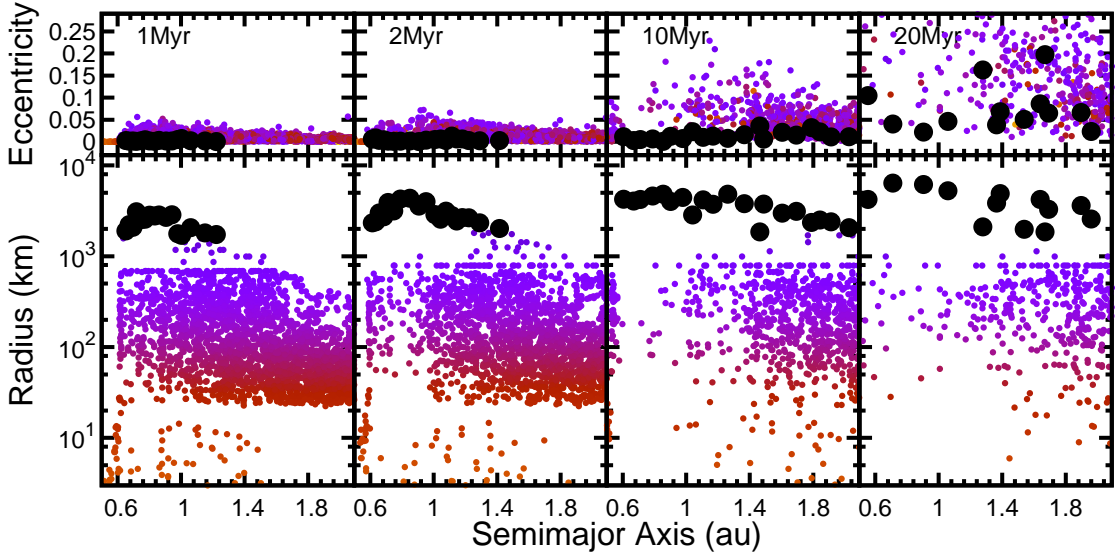}
\caption{ (Top) The orbital eccentricity and (Bottom) the radius (km)
  of the bodies are plotted as a function of their semi-major
  axis. The large closed circles are objects more massive than the Moon
  and their size is indicated in the bottom pane y-axis. The tracer particles have a color
  that corresponds to their size as found in the bottom pane. }
\label{fig:FourFrames}
\end{figure}

Here we present models of the growth and evolution of the inner Solar
System from km-sized planetesimals up to a final system of planets. We
include fragmentation of particles during collisions, interaction at
all sizes with a dissipating gaseous solar nebula, and the
gravitational interaction of all of the mass in the system. A Nominal
simulation and its initial conditions is shown in Figure
\ref{fig:FourFrames}. The growth is strongly inside-out, as found and
predicted by previous works
\citep{Weidenschilling:1997p23043,Kenyon:2006p11683,Minton:2014p18561,Carter:2015p19516}.
The localized area near 1~au reaches a 50\% mass ratio between
planetary embryos and planetesimals in 700~kyr (here objects more
massive than 0.012~Earth Masses, a lunar mass, are considered embryos
and less massive objects are planetesimals). This region is nearly
90\% planetary embryos by mass at 2~Myr and already a Mars-mass embryo
has grown. Meanwhile, at 2~au the same bi-modal state is reached in
approximately 18~Myr, and only reaches a point where 90\% of the mass
is in planetary embryos at 50~Myr and the largest body does not reach
a Mars mass due to grinding and drifting of planetesimals. The
planetesimal population is heavily depleted inside of 1~au by 2~Myr,
and is almost entirely gone by 10~Myr. Meanwhile at 1.5~au the
planetesimals are slightly depleted, but also dynamically excited (as
seen by their elevated eccentricity relative to those closer to
2~au). Despite being 5 e-folding times into the gas disk dissipation,
the system of embryos is still dynamically cold (see top panes in
Figure \ref{fig:FourFrames}). This has changed by 20~Myr, as the
embryos are in the midst of a instability with high eccentricities,
high velocity collisions and giant impacts. The excited embryos
naturally lead to the excited population of planetesimals.

\begin{figure}[!ht]
  \includegraphics[angle=0,width=3in]{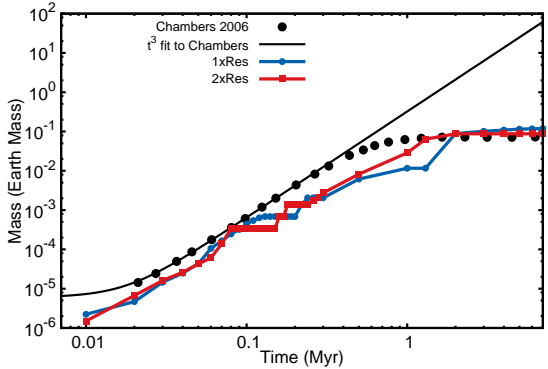} 
\caption{The growth of embryos at 1~au plotted as a function of mass (Earth
  masses) as a function of time (Myr). The black circles are taken
  from \citet{Chambers:2006p11697} Figure 1 with a fit to $t^3$
  through 200,000 years shown in the solid black line. The LIPAD
  Nominal simulation is shown in connected blue circles and the
  2$\times$ resolution Nominal simulation is shown with the red
  connected squares. }
\label{fig:ChambersTime}
\end{figure}

\begin{figure}[!ht]
\includegraphics[angle=0,width=3in]{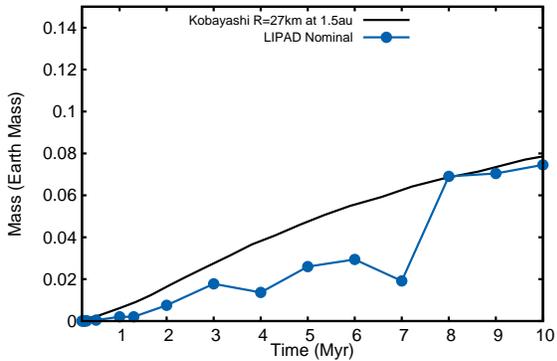} 
\caption{The growth of embryos out to 10~Myr as calculated in
  \citet{Kobayashi:2013p11732} for $R_\mathrm{init}=27$~km shown as
  mass (Earth masses) as a function of time (Myr). The connected blue
  circles are results from the Nominal LIPAD simulation at
  1.5~au. Note that the mass decreases at points, which is a result of some embryos moving in and out of the region 1.45--1.55~au that is analyzed to track growth in this region. }
\label{fig:KobayTime}
\end{figure}

The timescales for embryo growth can be directly compared to previous
analytical formulations. Embryo growth proceeds with roughly the
predicted scaling with time ($\sim$~$t^3$) for times up to a few
hundred thousand years, as found in the analysis of
\citep{Chambers:2006p11697}. This is recovered despite not considering
embryo atmospheres but including a full planetesimal size
distribution. The growth at 1~au is an especially close match beyond a few
hundred thousand years despite there also being a mis-match in initial
planetesimal size (these models start with 30~km as opposed to 10~km
used by \citet{Chambers:2006p11697}: see Figure
\ref{fig:ChambersTime}).  Similarly, embryo growth at 1.5au proceeds
largely as predicted by the analytical work found in
\citep{Kobayashi:2010p9661,Kobayashi:2013p11732}, despite a large
disparity in gas dissipation timescale (10Myr vs. 2Myr: see Figure
\ref{fig:KobayTime}).

\begin{figure}[!ht]
\includegraphics[angle=-90,width=3in]{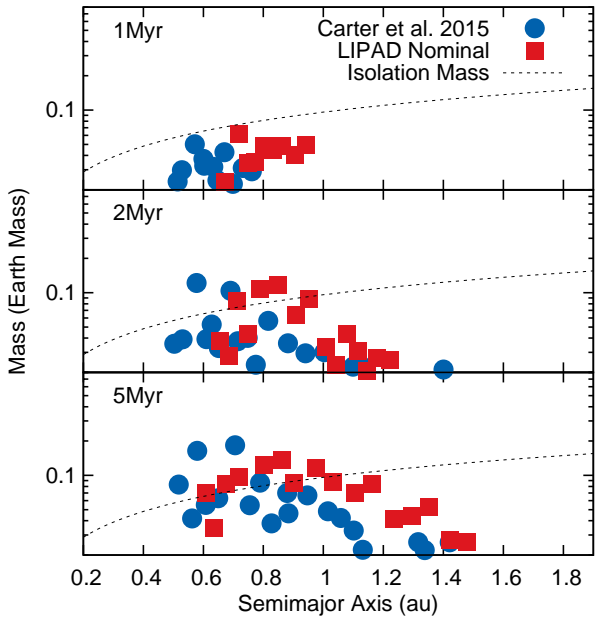}
\caption{Comparison of the LIPAD Nominal simulation from this work
  compared to that of \citet{Carter:2015p19516} at 1Myr, 2Myr and
  5Myr, with the isolation mass for 1$\times$MMSN plotted in each
  pane. The points from the ``LIPAD Nominal'' simulations represent
  embryos in the Nominal conditions plotting their mass as a
    function of their semimajor axis. The simulation from
  \citet{Carter:2015p19516} have an inner edge at 0.5~au (in contrast
  to the 0.7~au used in the Nominal LIPAD simulations), utilized an
  initial particle size distribution ranging from 196-1530~km, and
  rapid gas dissipation after 2~Myr.}
\label{fig:CarterTime}
\end{figure}

The growth of embryos in LIPAD can also be compared to the direct
$N$-body simulations utilizing gas affects and the EDACM collision
model (see Figure \ref{fig:CarterTime}). The two simulations being
compared used different inner edges of the disk, 0.5 and 0.7 au for
\citet{Carter:2015p19516} and this work respecively. They also used
different starting sizes for planetesimals, where
\citet{Carter:2015p19516} used an initial size distribution of
particles ranging between 196-1530~km meant to represent a disk
already in runaway growth, which is therefore at a more advanced stage
of growth than the Nominal LIPAD simulation initial conditions with a
unimodal size distribution $R_\mathrm{init}=30$~km.  Furthermore, the
direct $N$-body simulations of \citet{Carter:2015p19516} use very
different techniques, with very high resolution (100,000 particles)
and radius inflation to reduce runtime (where comparisons are made
between their scaled effective simulation times). Despite these
parameter and modeling differences, the growth and evolution of the
suite of embryos is qualitatively similar, showing strong inside-out
growth and embryos growing above expectations for isolation masses due
to embryo-embryo mergers and mass re-distribution in only a few
million years. These are important aspects to the current results
discussed in depth below.

Finally, the resulting planet systems can be compared with a
  plethora of previous works. This Nominal suite of simulations lost
  significant mass to collisional grinding (see Section \ref{massloss}
  below for more in depth discussion of mass loss). This mass lost
  drove the final mass in planets very low compared to the current
  Solar System, resulting in an average total mass of 1.44 Earth
  Masses (compared to 1.97 for the current Solar System). This
  resulted in too many planets (average of 9.5) on dynamically cold
  orbits (see Table \ref{bigtable}). There were nominally small Mars
  analogs, averaging masses of 0.25 Earth Mass, but in light of
  the total depletion of mass this does not represent a significant
  change in the Earth/Mars mass ratio. It also motivates the other
  parameter space that is explored in this work with more massive
  initial starting conditions.

To zero-th order these Nominal LIPAD simulations reproduce many of the
qualitative growth patterns and quantitative timescales previously
found.  While the inside-out growth has generally been predicted or
observed in numerical experiments before, it is commonly assumed, that
the disparity in timescales was less than the time of the transition
from oligarchic growth to chaotic growth or giant impacts (see
\citealt{Kenyon:2006p11683} who note that this is not always the case
in their simulations). The completion of oligarchic growth across the
entire disk provided convenient initial conditions for numerical
models that aim to model the final stages of planet growth and the
resultant planetary systems. Here, the outcomes of these models do not
support this assumption and warn future works away from assuming a
full disk of embryos amidst planetesimals in different places at
different times.

\subsection{Oligarchy moves outward like a wave}

Oligarch growth moves outward like a wave and is only valid in one
place in the disk at a time. This is more than just inside-out growth,
but rather a ``front'' (a term used by \citealt{Carter:2015p19516}) of
oligarchic growth progressing over time, such that only one small
region of the disk can meet any canonical oligarchy condition of mass
ratio between embryos and planetesimals.  The timescales for growth
vary as a function of semimajor axis due largely to differences in
orbital periods \citep{Chambers:2006p11697,Kobayashi:2010p9661}. One
analytic estimation for growth timescale from planetesimal accretion
alone, $\tau_\mathrm{grow,p}$ depends on semimajor axis, $a$, to the
$27/10$ power (see \citealt{Kobayashi:2013p11732} eq. 5), which alone
could account for a factor of three in growth timescales between 1au
and 1.5au.

\begin{figure}[!ht]
\includegraphics[angle=0,width=3in]{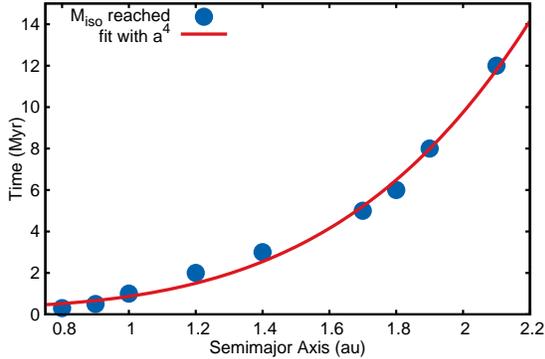}
\caption{The time at which the isolation mass is reached as a function
  of semimajor axis (au) for the nominal model (including nominal and
  high-resolution). The blue dots are data from both the standard and
  high-resolution simulations with a best-fit line drawn (red) with a
  dependence on $a^4$.}
\label{fig:Oligarch}
\end{figure}

In the nominal simulations (including the high-resolution case) and
the grid of twelve simulations exploring disk mass and disk lifetime,
the time at which isolation mass is reached can be described by $\sim
a^4$ to $\sim a^{4.2}$ (see Figure \ref{fig:Oligarch} showing the two
nominal simulations). This steeper relationship results in slower
relative growth at 1.5~au compared to 1~au, resulting in closer to a
factor of five in growth timescales (see Figure \ref{fig:Oligarch}).

\begin{figure}[!ht]
\includegraphics[angle=0,width=3in]{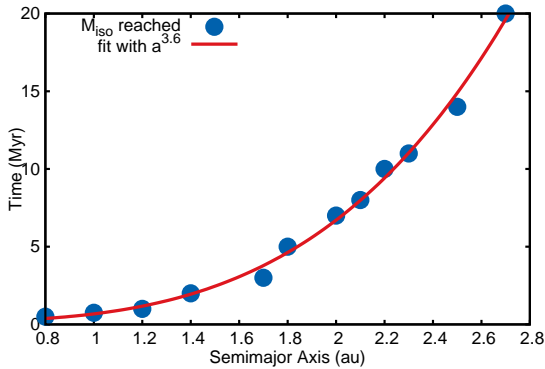}
\caption{The time at which the isolation mass is reached as a function
  of semimajor axis (au) for the nominal model with no collisional
  grinding. The blue dots are data from one simulation with a best-fit
  line drawn (red) with a dependence on $a^{3.6}$.}
\label{fig:OligarchNoColl}
\end{figure}

The likely reason for the stronger dependence on $a$ found in this
work compared to previous analytic efforts could be the more
substantial mass loss and drift due to collisional grinding. A test
simulation was run for the nominal scenarios where no collisional
grinding was allowed (collisions occurred and were tracked, but the
$Q^{\star}$ law was altered to make bodies unbreakable). The time to
reach isolation mass had a less steep dependence, best fit by
$a^{3.6}$ (see Figure \ref{fig:OligarchNoColl}). While this does not
entirely bridge the gap with the analytical estimates, there is a key
difference in modeling setups, where for any region in the analytical
models there is material and debris grinding and drifting from a more
distant semimajor axis. Here, the disks modeled have edges at 2.25~au
or 3.0~au, so there is likely some contribution to these differences
by the lack of material drifting in from the outside to partially
replace material that has drifted away to smaller semimajor axis.

\begin{figure}[!h]
\includegraphics[angle=-90,width=3in]{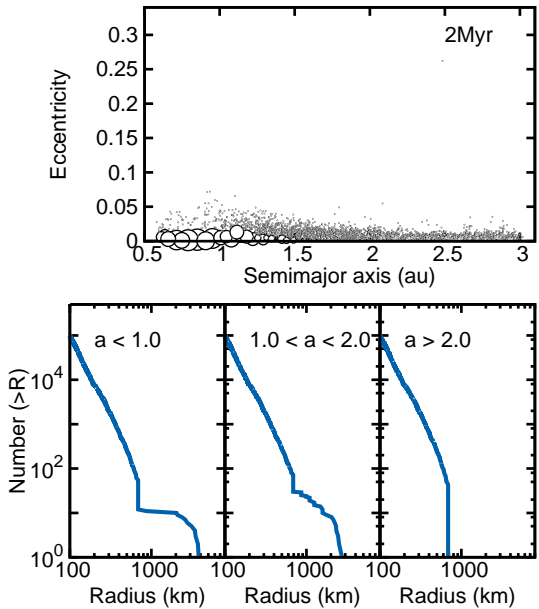}
\caption{The state of the inner-disk at 2Myr, (top) with the eccentricity of each body plotted as a function of their semimajor axis (au)
  with ``embryos'' shown in black scaled by their relative radius. The
  bottom three panes show (left) the size frequency distribution of
  particles for $a<1.0$~au, (middle) for $1.0<a<2.0$~au and (right)
  for $a>2$~au.}
\label{fig:ThreeSFD2Myr}
\end{figure}

This stronger inside-out growth means that for a minimal mass solar
nebula with $R_\mathrm{init}\sim30$~km, there is no time when the
entire disk is in runaway growth {\it or} oligarchic growth (see
Figure \ref{fig:ThreeSFD2Myr}). At 2~Myr the region inside 1~au is
dominated by embryos, making up 81\% of the mass, which is seen in the
break in the size frequency distribution (Figure
\ref{fig:ThreeSFD2Myr}, bottom left pane).  Between 1.0 and 2.0~au the
disk is 22\% embryos, and there is no break in the size distribution
pointing to most of the region still growing by runaway
growth. Finally, beyond 2~au, there are only planetesimals the largest
of which has grown to 787~km and runaway growth is moving along
slowly.

\begin{figure}[!h]
\includegraphics[angle=-90,width=3in]{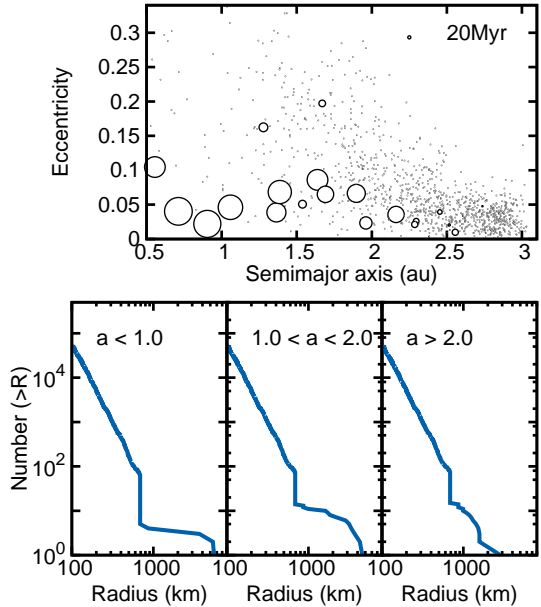}
\caption{The state of the inner-disk at 20Myr, (top) with the eccentricity of each body plotted as a function of their semimajor axis (au)
  with ``embryos'' shown in black scaled by their relative radius. The
  bottom three panes show (left) the size frequency distribution of
  particles for $a<1.0$~au, (middle) for $1.0<a<2.0$~au and (right)
  for $a>2$~au. }
\label{fig:ThreeSFD20Myr}
\end{figure}

The timescales across the inner disk are so vastly different that by
20~Myr the region inside of 1~au is experiencing chaotic growth and
giant impacts, but beyond 2~au the disk has built its first few
embryos (see Figure \ref{fig:ThreeSFD20Myr}). Despite the successes of
analytical models in calculating embryo growth some of the disk
entering a chaotic phase challenges any analytical description of
final planet growth owing to the chaotic nature of embryo-embryo
collisions and scattering, This same affect also challenges numerical
techniques that can't handle simultaneous modeling of such different
regimes of growth.

\subsection{Mass loss due to grinding has a strong dependence on semimajor axis} \label{massloss}

Different embryos that have similar mass but are growing in different
regions of the disk, experience different growth conditions over their
lifetime due to differences in mass loss and growth timescales
relative to the gas disk lifetime.  At 1~au a bi-modal mass
distribution is reached after only 700~kyr, which is $\sim$1/2 of the
gas disk's e-folding dissipation time. This is in stark contrast to
18~Myr at 2~au, which is $\sim$9 e-folding times of the gas disk
dissipation timescale. The enormous difference in gas quantities
results in more effective stirring by the embryos at greater distance,
where the RMS eccentricity of planetsimals is 0.014 at 700~kyr at
1~au, while at 18~Myr at 2~au it is 0.1 (these translate to average
collision velocities of $\sim$0.46~km/s and 2.3~km/s respectively).

\begin{figure}[h]
\includegraphics[angle=-90,width=3.5in]{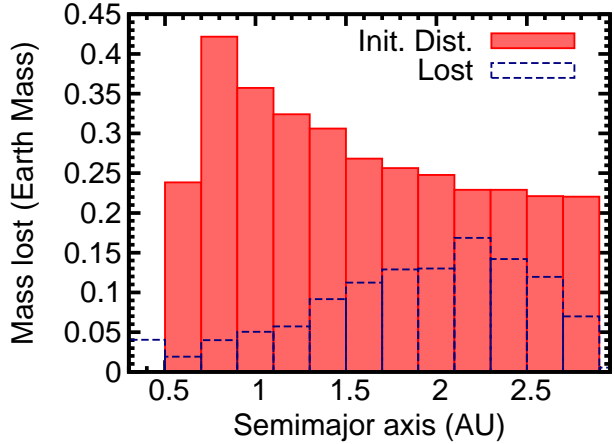}
\caption{Mass distribution at the outset of the Nominal simulation (red filled
  histogram) and lost to grinding during the simulation (blue outline)
  as a function of semimajor axis for the nominal simulation during
  the first 25~Myr.}
\label{fig:Money}
\end{figure}

This disparity leads to a strong preferential mass loss at greater
distances (Figure \ref{fig:Money}), owing to the higher eccentricities
(Figure \ref{fig:FourFrames}) and thus more violent and disruptive
impacts. The location of maximum mass loss tracks outward with
  the oligarchic wave, so that at very early times (0-5~Myr), most of
  the collisional grinding is happening around 1~au - where the first
  embryos are growing. As embyros start appearing further out, mass
  loss follows, as at smaller distances the larger embyros have
  heavily depleted their local planetesimal population and at further
  distances there are no embryos around yet to stir the
  planetesimals.

\begin{figure}[h]
\includegraphics[angle=-90,width=3.5in]{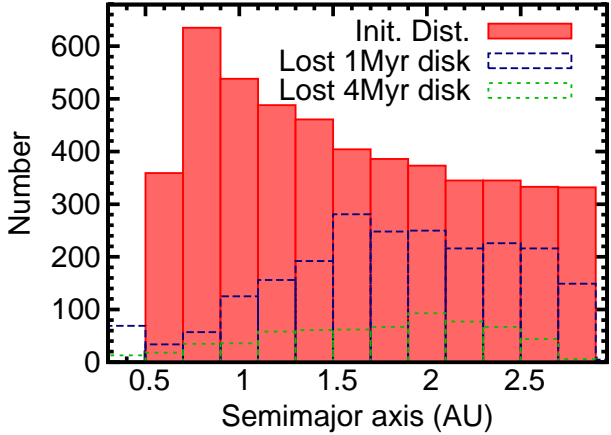}
\caption{Mass distribution at the outset of the simulation (red filled
  histogram) and number of particles lost in a disk with a 1Myr
  lifetime (blue outline) and lost in a disk with a 4Myr lifetime
  (green) as a function of semimajor axis for the nominal simulation
  during the first 25~Myr.}
\label{fig:DustGasLife}
\end{figure}

The magnitude of mass loss to collisional grinding is known to also be
a function of the initial planetesimal size (as is the growth
timescale, which here is relevant since the gas disk dissipation
timescale is not changing; see \citealt{Kobayashi:2013p11732}).  Here,
when the initial planetesimal radius is 3~km or 100~km (instead of
30~km in the Nominal case) there is more and less collisional grinding
respectively, but varying by no more than a factor of two. In the case
of $R_\mathrm{init}=$100~km, with the least mass loss, more then 20\%
of initial mass at 1.5~au was lost to disruptive impacts. Similarly,
extending or decreasing the gas disk lifetime can alter the amount and
profile of mass loss. For a simple test over 25~Myr, with a gas disk
lifetime or 1~Myr and 4~Myr, the shorter disk lifetime allows earlier
excitation of the system and more collisional grinding and mass loss
(Figure \ref{fig:DustGasLife}).

This result leads naturally to ask how this substantial mass loss
affects the final planetary systems since it appears to offer a path
to depart from the initial surface density distribution of solids.

\subsection{Can grinding account for Earth/Mars mass ratio?}

As found above, mass loss due to collisional fragmentation has a
strong dependence on semi-major axis, with mass loss maximized at and
beyond where oligarch growth has reached by the time that the gas-disk
has dissipated $\sim$5 e-folding times. At large heliocentric
distances the lack of damping due to the gas means that accretion
efficiencies are lower and collisional fragmentation is increased
amidst planetary embryos that are stirring the local population.  For
a suite of four low-resolution nominal case simulations (nominal
initial conditions, but only 2000 tracer particles), the grinding
causes significant mass loss around 1.5~au, and production of a
reasonable Mars-analogs ($\sim 0.25 M_\oplus$). However, the total
amount of grinding affected the masses of the Earth and Venus analogs,
leaving only 1.44$M_\oplus$ of total mass in planets after 115~Myr
(see Table \ref{bigtable}). This mass loss resulted in many planets
remaining on dynamically cold orbits (see N$_\mathrm{planets}$ and AMD
in Table \ref{bigtable}). While longer run-times may have found
eventual reduction in the total number of planets, the total planet
mass was so low that the runs were not continued beyond 115~Myr.

The Nominal results motivate an exploration of disk mass and gas disk
dissipation timescale. This is similar approach to that of
\citet{Kobayashi:2013p11732} who sought a combination of initial
planetesimal radius and total disk mass to accomodate the mass of
embryos at 1.5~au and the short accretion timescale of Mars. Here, the
relation is that with increasing disk lifetime the point in the disk
where grinding is maximized moves outward as Oligarch growth will move
further due to the longer time amidst gas. But the growth rate
increases with increasing total disk mass, as does the expected
isolation masses of embryos. By testing combinations of
1$\times$,$1.5\times$,2$\times$MMSN paired with gas disk timescales of
3~Myr, 2~Myr and 1~Myr we aim to balance increased solid material with
less or more rapid growth timescales and less or more total
grinding. For each set of parameters four low-resolution simulations
were performed.

This grid of simulations did not find a sweet spot of a depressed
Mars mass and $\sim 2 M_\oplus$ of planets (see Table
\ref{bigtable}). The set of parameters that was closest to $\sim 2
M_\oplus$ of total mass of planets was for a 1.5x~MMSN and 2~Myr gas
lifetime, with an average of 2.22~$M_\oplus$. However, the Mars
analogs were large and averaged 0.6~$M_\oplus$. The Mars analogs are
lower for the 1x~MMSN 3~Myr gas cases, 0.43~$M_\oplus$, but the total
mass in all planets was only 1.62~$M_\oplus$.

For each simulation in the grid of runs, the system was analyzed at
115~Myr, tracking the total mass of planets M$_\mathrm{tot}$, where a
planet is defined to be 1/30~th of an Earth Mass. Also tracked are the
mass of Mars analogs (M$_\mathrm{M}$; if $1.2<a<2.0$) and their
semimajor axis at completion (a$_\mathrm{M}$), and the system Angular
Momentum Deficit (AMD) and the Radial Mass Concentration (RMC). 
The AMD measures the dynamical excitement of a system of planets
relative to the same system with zero eccentricity and incliation
\citep{Laskar:1997p23044,Chambers:2001p7618}, where the current
terrestrial planets value is 0.0014, and is defined as

\begin{equation}
AMD = \frac{\sum_j M_j \sqrt{a_j}(1-\cos(i_j)\sqrt{1-e_j^2})}{\sum_j M_j \sqrt{a_j}} 
\label{eq2}
\end{equation}
where $M_j$, $a_j$, $i_j$, and $e_j$ are the mass, semimajor axis,
inclination, and eccentricity of planet $j$. 

The RMC is a metric that increases with concentration of planetary
mass, and decreases with widespread systems with distributed planetary
mass, where the current solar system value is 89.9 and its calculated by 
\citep{Chambers:2001p7618}.
\begin{equation}
RMC = {\rm max}\Big(\frac{\sum M_j}{\sum M_j [\log_{10}(a/a_j)]^2}\Big)
\label{eq3}
\end{equation}
\noindent where $M_j$ and $a_j$ are again the mass and semimajor axis
of planet $j$, and $a$ ranges from 0.1 au to 2 au. The RMC is
particularly sensitive to the small Mars problem, where results with
the typically over-sized Mars-analogs of 5-10$\times$ the size of Mars
produce RMC values between 30--50 \citep{Raymond:2009p11530}.

\begin{figure}[h]
\includegraphics[angle=0,width=7.2in]{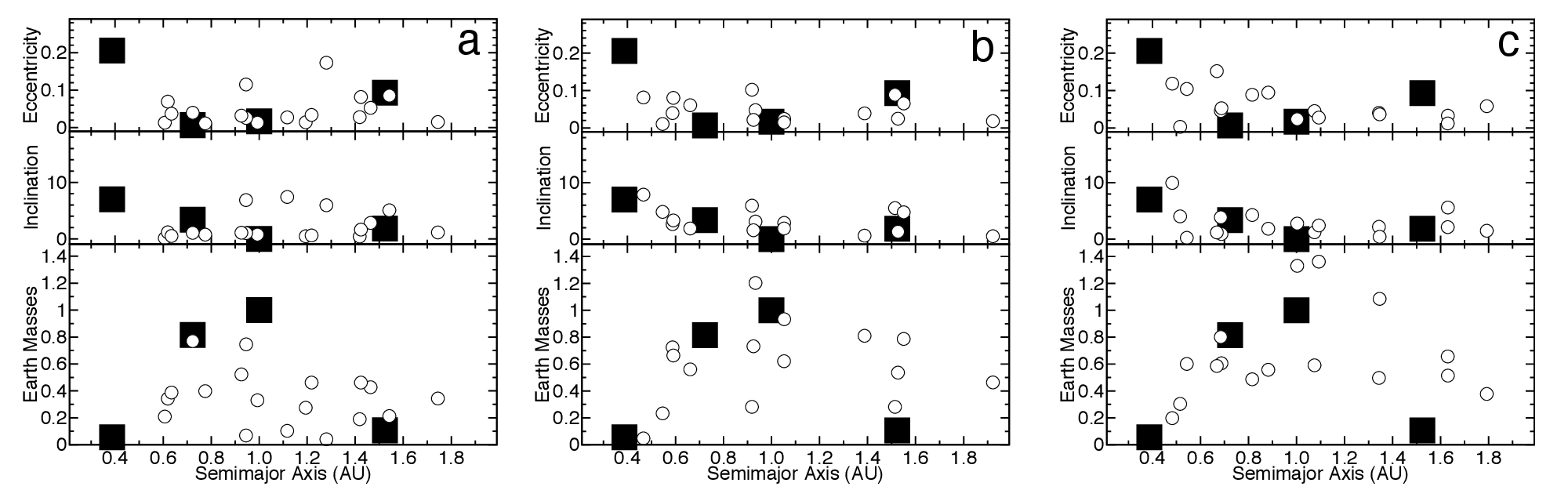}
\caption{The final planets remaining in the four simulations for (a) 1$\times$MMSN with a 3~Myr gas disk dissipation timescale. The produced planets are the open circles and the actual planets are the filled squares, and they are both plotted as (top) their orbital eccentricity (middle) orbital inclination in degrees and (bottom) their mass in Earth masses as a function of their semi-major axes in Astronominal Units (au). Pane (b) shows the same data for four simulations with 1.5$\times$MMSN with a 2~Myr gas disk dissipation timescale, and (c) shows four simulations of a 2$\times$MMSN with a 1~Myr gas disk dissipation timescale. }
\label{fig:123MMSN}
\end{figure}

The planetary systems created in the partial grid of parameters space,
from 1-2$\times$MMSN and 1-3~Myr gas disk dissipation time, produced
roughly consistent systems within each set of parameters (see Figure
\ref{fig:123MMSN} and Table \ref{bigtable}). The 1$\times$MMSN case has too little mass remaining, averaging
1.62~$M_\oplus$ due to mass lost during collisional grinding. The mass
of Mars is the lowest of the three cases, averaging 0.43~$M_\oplus$,
but simply decreased proportional to the total deficiency of mass in
the planetary system. The deficiency of total mass also results in low
mass Earths (Figure \ref{fig:123MMSN}a), and the largest number of total
planets, 4.5, on average.

The 2$\times$MMSN overshot the total mass, averaging 2.63~$M_\oplus$
per system of planets, and also large Mars analogs averaging
0.69~$M_\oplus$. This is expressed in the handful of
$\sim$1.3~$M_\oplus$ planets formed in these simulations (Figure
\ref{fig:123MMSN}b). Finally, the 1.5$\times$MMSN initial disk with a
2~Myr gas disk lifetime provided the best match to the total mass of
the suite of final planets, averaging 2.22~$M_\oplus$, but still far
overshooting the mass of Mars analogs with an average of
0.6~$M_\oplus$.

All of the parameter suites produced similar AMD and RMC values. The
similarity of the RMC shows that the mass of Mars is increasing or
decreasing roughly in proportion to the total mass of planets for each
set of parameters. The values of RMC are roughly in line with those
found for numerous previous simulation scenarios including giant
planet excitation \citep{Raymond:2009p11530}, and well below those
found by the scenarios that utilize an annulus that have found good
matches for this particular metric.
\citep{Hansen:2009p8802,Jacobson:2014p18340,Brasser:2016}.

A large suite of classical simulations that varied primarily the giant
planet orbit conditions produced final AMD values between 1-10$\times$
the current value, so between 0.0018 and 0.018
\citep{Raymond:2009p11530}.  A large suite of annulus, or Grand Tack,
simulations found that AMD values for final planets would vary between
$10^{-4}$ and 0.01 for a range of tested initial embryo mass and
embryo size \citep{Jacobson:2014p18340}. The averages found here, 0.0030 -- 0.0046, or
$\sim$2--3$\times$ the current value represent no clear improvement
over the outcomes described in \citet{Raymond:2009p11530}, and are
possibly slightly worse than that found in the Grand Tack cases tested
in \citet{Jacobson:2014p18340}.

The RMC values for \citet{Raymond:2009p11530} tests range between
0.2--0.8$\times$ the current value. Naturally
\citet{Jacobson:2014p18340}, doing annulus tests that have a long
record of producing good Earth/Mars mass ratios, produced excellent
RMC values 0.5--1.1 times the present RMC. Meanwhile the
simulations presented here look more similar to the full-disk tests of
\citet{Raymond:2009p11530} and consistently produce RMC values between
30--60, or 0.3--0.75 times the current RMC value, suggesting Mars-analogs
that are too massive.

There is no parameter set tested here, of gas disk lifetime and disk
mass, using 30~km initial planetesimal radii, that can satisfy both a
small Mars and supply the total mass required by the four terrestrial
planets.

\begin{table*}[p]
\begin{tabular}[h]{l|ccccccccc}
 R$_\mathrm{init}$ (run \#) & N$_\mathrm{planets}$ & M$_\mathrm{tot}$ & AMD  & RMC & M$_\mathrm{M}$ (M$_\oplus$)&a$_\mathrm{M}$ (au) \\
\hline
1xMMSN 2Myr gas (1) & 19 &1.39 &0.0001 & 42.82 & 0.17 &1.51\\
1xMMSN 2Myr gas (2) & 8 & 1.40 & 0.0006 & 48.75 & 0.19 & 1.24\\
1xMMSN 2Myr gas (3) & 5 & 1.43 & 0.0008 & 40.73 & 0.37 & 1.37 \\
1xMMSN 2Myr gas (4) & 6 & 1.55 & 0.0007 & 43.17 & 0.27 & 1.22 \\
\hline
Average & 9.5 & 1.44 & 0.0006 & 43.87 & 0.25 & 1.34 \\
\hline
\hline
2xMMSN 1Myr gas (1) & 3 & 2.64 & 0.0029 & 59.13 &  0.51 & 1.63\\
2xMMSN 1Myr gas (2) & 5 & 2.55 & 0.0031 & 32.40 &  0.50 & 1.34\\
2xMMSN 1Myr gas (3) & 4 & 2.55 & 0.0019 & 43.49 &  1.08 & 1.34\\
2xMMSN 1Myr gas (4) & 4 & 2.79 & 0.0039 & 40.52 &  0.66 & 1.63\\
\hline
Average 2x 1Myr gas   & 4 & 2.63 & 0.0030 & 43.89 &  0.69 & 1.49\\
\hline 
\hline
1.5xMMSN 2Myr gas (1) & 3 & 1.94 & 0.0027    & 44.32  & 0.28 &1.52\\
1.5xMMSN 2Myr gas (2) & 3 & 1.97 & 0.0021    & 56.02  & 0.54 &1.53\\
1.5xMMSN 2Myr gas (3) & 4 & 2.35 & 0.0125    & 33.07  & 0.79 & 1.59\\
1.5xMMSN 2Myr gas (4) & 5 & 2.61 & 0.0011    & 35.52  & 0.81 & 1.39\\
\hline
Average 1.5x 2Myr gas   & 3.5&2.22 & 0.0046    & 42.23  & 0.6  & 1.5\\
\hline
\hline
1xMMSN 3Myr gas (1) & 6   & 1.75 & 0.0024    &43.97   & 0.34 & 1.74\\
1xMMSN 3Myr gas (2) & 4   & 1.60 & 0.0063    &47.22   & 0.47 & 1.50\\
1xMMSN 3Myr gas (3) & 4   & 1.44 & 0.0036    &53.11   & 0.46 & 1.43\\
1xMMSN 3Myr gas (4) & 4   & 1.68 & 0.0030    &49.05   & 0.46 & 1.22\\
\hline
Average 1x 3Myr gas   &4.5  & 1.62 & 0.0038    &48.34   & 0.43 & 1.47\\
\hline
Current Solar System & 4  & 1.97 & 0.0014 & 89.9   & 0.11 & 1.52\\
\hline
\hline
$R_\mathrm{init}=5$~km 2xMMSN 2~Myr gas (1) & 4 & 2.41 & 0.0019 & 41.5 & 0.60 & 1.41\\
$R_\mathrm{init}=5$~km 2xMMSN 2~Myr gas (2) & 4 & 2.47 & 0.0020 & 35.5 & 0.61 & 1.73\\
$R_\mathrm{init}=5$~km 2xMMSN 2~Myr gas (3) & 2 & 2.38 & 0.0032 & 40.8 & --   &  -- \\
$R_\mathrm{init}=5$~km 2xMMSN 2~Myr gas (4) & 4 & 2.69 & 0.0026 & 37.0 & 0.28 & 1.71\\
\hline
Average 5~km, 2xMMSN, 2~Myr gas & 3.5 & 2.49 &0.0024 & 38.7 & 0.50 & 1.61\\
\hline
\hline
\end{tabular}
\caption{Total number N$_\mathrm{planets}$ and mass M$_\mathrm{tot}$
  at 115~Myr, and their AMD and RMC values.  Mass of Mars analogs
  (M$_\mathrm{M}$; if $1.2<a<2.0$) and their semimajor axis at
  completion (a$_\mathrm{M}$). \label{bigtable} }
\end{table*}

\subsection{Growing Mars from Small Planetesimals in a Massive Disk}

\citet{Kobayashi:2013p11732} propose that the combined constraints of
the chronology of Mars accretion ($\sim$few million years) and its small
mass relative Earth can be accounted for by the growth from a massive
(a few times MMSN) disk of small planetesimals (less than 10~km). The
timescales to reach oligarchic growth are faster for the smaller initial
planetesimal size, and the increased disk mass makes up for loss to
collisional grinding. However, it is less clear what will happen at
1~au - whether a suitable set of Earth and Venus analogs will form -
and how many other Mars-mass embryos could grow nearby or beyond any
that grow at 1.5~au.

\begin{figure}[h]
\includegraphics[angle=0,width=4in]{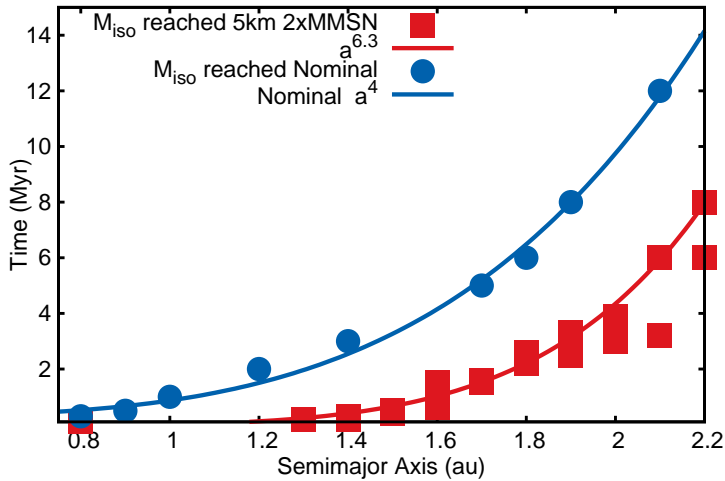}
\caption{The time at which the isolation mass is reached as a function of semimajor axis (au) for the nominal model with initial planetesimals with radius of 5~km.}
\label{fig:Kob5km}
\end{figure}

\begin{figure}[h]
\includegraphics[angle=-90,width=4in]{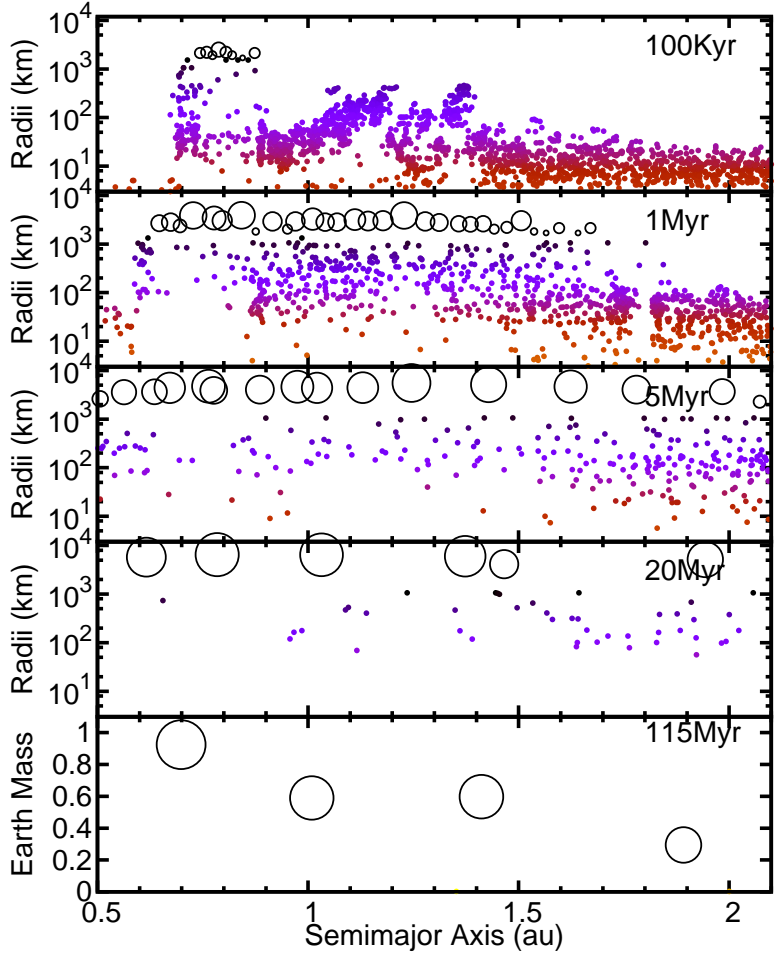}
\caption{The distribution of particle sizes at 100~kyr, 1~Myr, 5~Myr and 20~Myr plotted as a function of Radius (km) versus semimajor axis (au). The bottom frame shows the mass of the final bodies in the system plotted as a function of semimajor axis. }
\label{fig:KobFrames}
\end{figure}

We performed a suite of simulations with $R_\mathrm{init}=5$~km and a
2$\times$MMSN disk with a 2~Myr gas disk lifetime. The growth is
indeed substantially faster, reaching isolation masses roughly twice
as fast for the Nominal scenario (see Figures \ref{fig:Kob5km}
and \ref{fig:KobFrames}). In only $\sim$1~Myr embryos have grown out
beyond 1.5~au, and by 5~Myr they stretch across the entire disk and
dominate the total mass of the system. 

The system at 5~Myr is nearly an ideal description of a bi-modal mass
distribution - or the simple clean idea of oligarchic growth
conditions so frequently assumed in models. It is not entirely so
simple; between 0.7--1.0~au the embryos account for 94\% of the mass,
but only 53\% between 1.7--2.0~au. The planetesimal distribution is
not smooth, rather it shows a less extreme inside-out depletion as
found in the Nominal simulations discussed above.

\begin{figure}[h]
\includegraphics[angle=-90,width=3.2in]{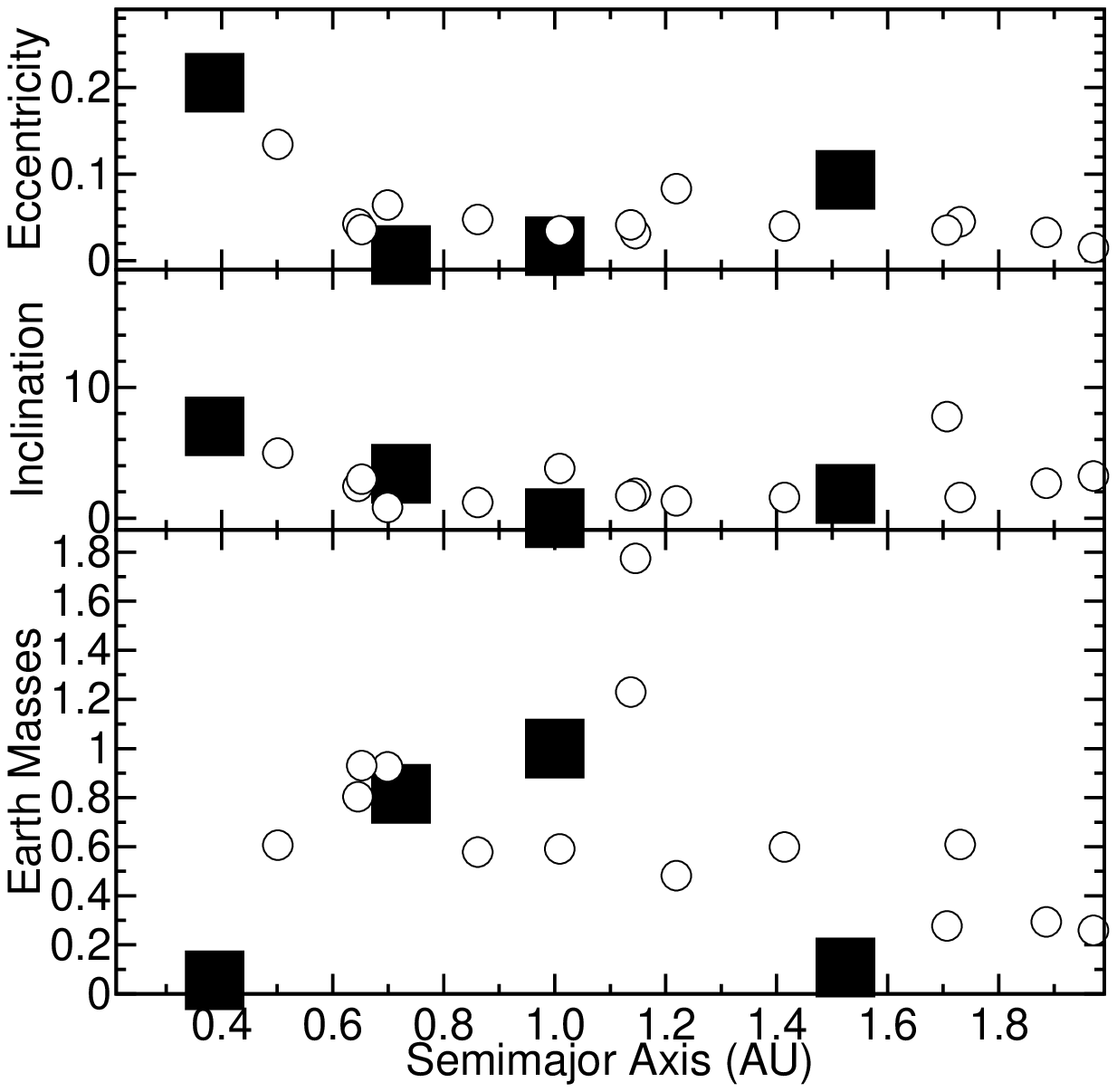}
\caption{The final planets remaining in the two simulations of a 2$\times$MMSN with a 2~Myr gas disk dissipation timescale with $R_\mathrm{init}=$5~km. The produced planets are the open circles and the actual planets are the filled squares, and they are both plotted as (top) their orbital eccentricity (middle) orbital inclination in degrees and (bottom) their mass in Earth masses as a function of their semi-major axes in Astronominal Units (au).}
\label{fig:PlanetsKob}
\end{figure}

The final planets meanwhile suggest that not enough total mass was
lost due to collisional grinding (Figure \ref{fig:PlanetsKob}). The
four simulations had similar final masses of planets all over
2.49~$M_\oplus$. The Mars-analog on average were only two to three
times larger than Mars, but that was balanced by the simulation that
only produce 2 planets with no Mars-analog. Meanwhile, of the few
Earth-analogs formed at 1~au one was also 60\% of an Earth-mass, while
another was 1.77~$M_\oplus$. Without a more complete suite of
simualtions its hard to entirely discount this scenario, but these
first results suggest that while the accretion timescale and embryo
mass at 1.5~au might be matched with the collateral outcome of too
much mass at 1~au resulting in too few and/or too massive planets.

\subsection{Two generations of planets}

We find that embryos resulting from Oligarchic growth remain stable
even after the vast majority of planetesimal mass are gone, where at
10~Myr the region inside of 1~au is 91\% embryo by mass and is 94\% by
15~Myr. This stability results in a quiescent period lasting
$\gtrsim$10~Myr prior to the onset of the giant impact stage of planet
formation that is characterized by violent collisions between the
embryos.

\begin{figure}[h]
\includegraphics[angle=-90,width=3.5in]{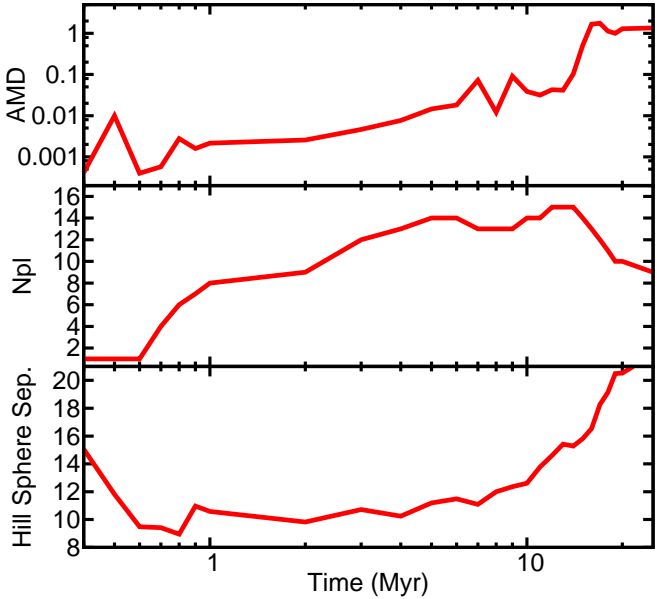}
\caption{(Top) The system Angular Momentum Deficit as a function of
  the current system's AMD, (Middle) the number of planets and
  (Bottom) the average relative Hill Sphere spacing between embryos,
  plotted as a function of time (Myr).}
\label{fig:Hill}
\end{figure}

This dynamically cold and quasi-stable phase lasts longer than
expected for similar systems without any gas affects.  The spacing of
the embryos found here, 10-12 mutual Hill Spheres, are the same as
typically found in numerical and theoretical models (Figure
\ref{fig:Hill})
\citep{Kokubo:1998p9706,Chambers:2006p11697,Kobayashi:2010p9661}.
Absent dynamical friction from planetesimals, and the affects of
damping from gas, this spacing is not stable on long timescales for an
isolated system of embryos, where 10 bodies with non-uniform, but
average, spacing of 10 mutual Hill Sphere and non-uniform masses should
only be stable for $\sim$1~Myr \citep{Chambers:1996p18902}. Owing to
the presence of the dissipating gaseous solar nebula, the system of
embryos remains dynamically cold (low eccentricity and inclination)
and quasi-stable.  The evolution of dynamical excitement is seen in
the history of the system's Angular Momentum Deficit (AMD), a common
metric to track a systems divergence from circular and uninclined
orbits \citep{Laskar:1997p23044}, where here a very slow increase in AMD
is seen from 1-10~Myr. Only after 10~Myr does the AMD rapidly increase
by over an order of magnitude in a few Myr (Figure \ref{fig:Hill}).
The sharp increase of the AMD reveals the dynamically cold suite of
embryos transitioning into the giant impact stage of planet formation.

These dynamics are dominated by the effects of the gas disk. The
dynamical stability of multi-planet systems has been found to change
with only 0.1\% of the starting MMSN of available gas
\citep{Iwasaki:2001p21107,Kominami:2002p21136}, which is similar to
the amount left in these simulations after $\sim$5 e-folding times,
$\sim$10~Myr, when an instability is typically first encountered.
Similar dynamics is found in other recent studies utilizing an
exponentially dissipating gas disk
\citep{Walsh:2016p23062,Levison:2015p20212}, where two very different
pathways to a suite of embryos, an annulus or pebble accretion, both
result in similar quasi-stability of the initial embryos.  The two
different timescales, a few Myr for building the first suite of
embryos and a few 10's Myr for completing the accretion of the planets
is similar to the expected accretion times for Mars and Earth
respectively \citep{Dauphas:2011p19768,Kleine:2009p9784}, which
provides some support for the idea that Mars is essentially an embryo
that sat out the final stage of growth.

The evolution of the system can be viewed as two generations
of planets, the first grown in a few Myr and only after
$\gtrsim$10~Myr will the final set of planets start to be built. This
is something not typically captured in models that begin at the
Oligarchic growth stage in the {\it absence} of gas.

\section{Rough Prescription for Generating Initial Conditions}

One big takeaway from the above discussion is that planet formation
initial conditions are vastly different than the typically assumed
Moon to Mars sized embryos amidst a sea of planetesimals. This
condition is never reached in any of our simulations, and here we
endeavor to provide sufficient descriptions to enable others to build
a set of initial conditions that match the growth described here.

\begin{figure}[h]
\includegraphics[angle=0,width=5in]{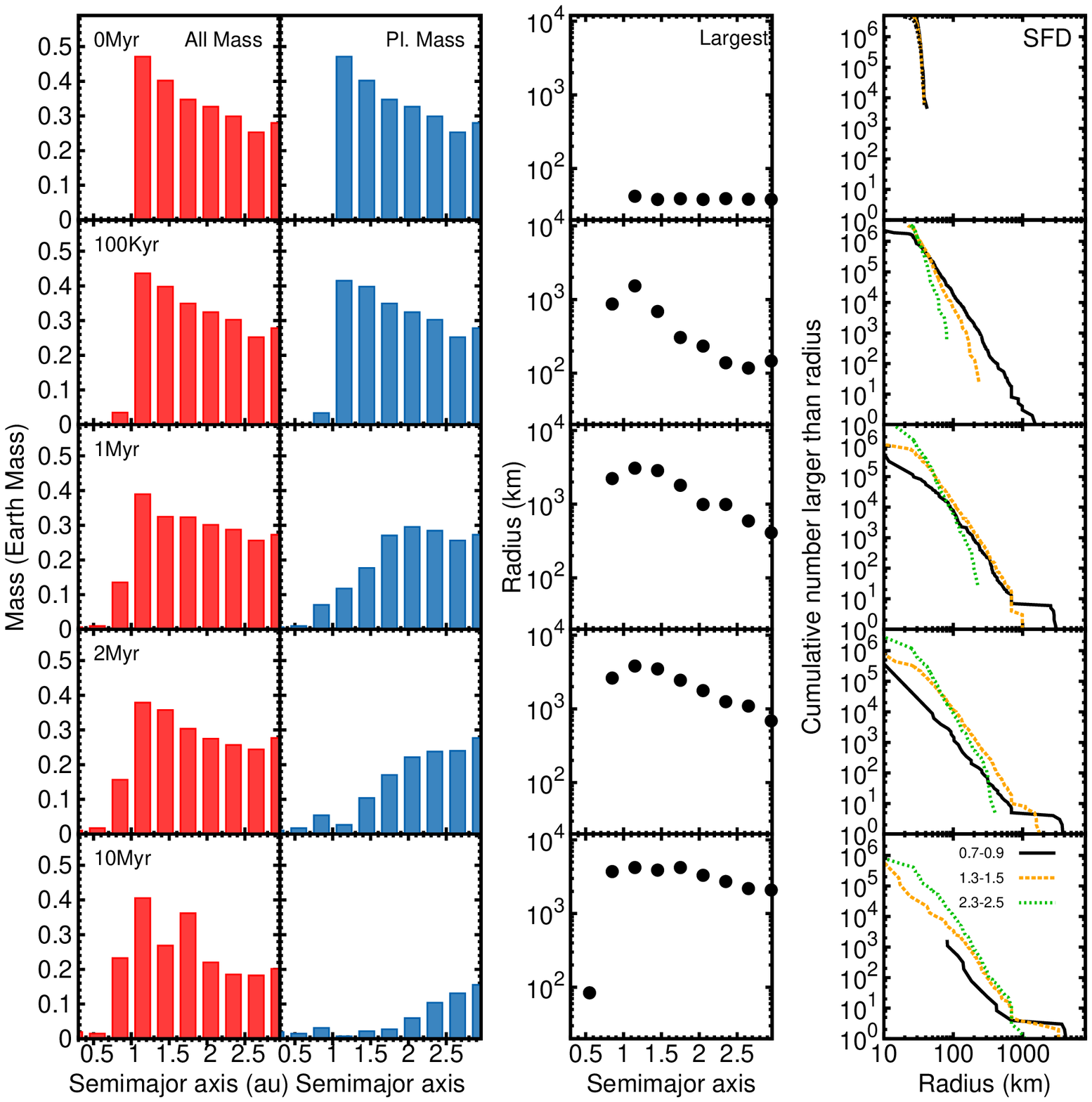}
\caption{The evolution of the Nominal simulation over the first 10~Myr
  showing the (left) total mass and (center) planetesimal mass as a
  function of semimajor axis (au). The third column of panes show the
  size of the largest body in kilometers as a function of semimajor
  axis and times. Finally, the fourth column shows the size frequency
  distribution in three different semimajor axis ranges: in black
  solid line is 0.7--0.9~au, in orange dashed line is 1.3--1.5~au and
  the green dotted line is 2.3--2.5~au.}
\label{fig:SD}
\end{figure}

At any given time knowledge of the regional size frequnecy
distribution (SFD) and orbital properties at each semimajor axis should
encapsulate the key pieces of information. An example of this also
displays the breakdown between all mass and planetesimal mass (see
Figure \ref{fig:SD}), as well as the largest body in each semimajor
axis bin and particle size-frequency distributions in three distinct
locations in the disk. The relationship between total mass and
planetesimal mass will allow for determination of embryo mass, and the
size of the largest body will provide, at least roughly, the dynamical
excitment of the local planetesimal population. The complete set of
data files showing orbit and size properties of every body are
available over entire simulations online (at
\url{www.boulder.swri.edu/~kwalsh/LIPAD.html}).

\section{Discussion and Conclusions}

Terrestrial planet formation is a deeply studied problem, but despite
the attention that it garners, effects studied for over a decade are
regularly not taken into account in modern discussions and
simulations. Furthermore we find here that some of the affects are
exaggerated when the dynamics of collisional evolution and the
dissipating gas disk are utilized for the entire growth from
planetesimals to planets.

The three questions posed in Section 1.1.3 questioned where and when
one can find oligarchy in a disk and the implications for the
planetesimal population. The Nominal scenario (1$\times$MMSN, 2~Myr
gas disk dissipation timescale and 30~km~$R_\mathrm{init}$) finds that
the growth timescale relative to the gas disk dissipation timescale
allow for rapid progression of oligarchy inside 1~au leading to rapid
depletion of planetesimals and rapid construction of a suite of
dynamically cold and stable embryos. It takes another 10~Myr years for
this state to reach the outer edges of the tested regions, by which
time the depleted gas disk allows for much more planetesimal loss to
collisional grinding. In this case chaotic growth starts inside of
1~au before oligarchic growth has reached beyond 2~au. Enough mass is lost to leave behind insufficient mass to build a suite of terrestrial planets.

In the simulations designed to test the $R_\mathrm{init}$=5~km and
2$\times$MMSN the growth timescale was significant faster than the gas
dissipation timescale and a full suite of disk-wide embryos formed
before the giant impact stage commenced. These simulations still had
significantly more depletion of planetesimals in the inner disk
compared to the outer.

The big takeaway is that exploring a modest grid of disk mass and gas
disk dissipation timescales there was no clear sweetspot where
collisional grinding could account for the mass difference between
Earth and Mars. Extending this to the proposed solution of
\citet{Kobayashi:2013p11732} of $\sim$5~km initial planetesimals also
did not seem to address the issue of the Earth/Mars mass ratio
(although it did decidedly speed up growth of Mars-sized embryos, as
predicted in that work). While there is significant parameter space
yet to explore there is no guarantee that there is a simple answer to
this issue based on collisional evolution, despite some clear trends,
basic physics and the attractive simplicity of it. Furthermore,
collisional grinding may not need to solve this problem alone, rather
a combination of different growth patterns (e.g pebble accretion; see
\citealt{Levison:2012p12338}) or larger solar system evolution (e.g an
early giant planet instability; see \citealt{Clement:2018}) may also
contribute to changing the Earth/Mars mass ratio.

Another way to view these findings are that despite the discussed
  differences in collisional mass loss and long-lasting suite of
  stable embryos, the planetetary systems are not wholly different
  than previous works. Once the mass of the disk was increased to 1.5
  or 2.0$\times$~MMSN, and the final total planet mass was close to 2.0
  Earth masses, the typical metrics for modeling the terrestrial
  planets showed values similar to nominal cases from
  \citet{Raymond:2009p6769}. While differences will still lurk in the
  outcomes for the growth timescales, number and nature of giant
  impacts and the structure of the remaining asteroid belt, it shows
  generally that the starting mass distribution (and its profile) will
  govern the final mass distribution in the terrestrial planets.

Future work should focus on departing from the simple assumption about
an exponential decay of the gas disk, which may be biasing results
away from any sharp discontinuities in embryo size or growth timescale
at specific distances. Similarly, departing from simple
  monodisperse distribution of initial planetesimal sizes could play
  an important role in changing the outcome of the simulations. 
Also, a change in the initial surface density profile of the disk
could also contribute to an increase in the Earth/Mars mass
ratio. Finally, these results are compiled from combining suites of
four simulations per parameter set and may not provide robust
statistical views of the possible outcomes.

The test of \citet{Kobayashi:2013p11732} also highlighted that in a
regime where the growth timescale is substantially shorter than the
the gas disk dissipation timescale, the disk did nearly reach a
bi-modal mass distribution (although the planetesimal population did
not follow the original surface density profile). Although not tested
here, one can speculate that substantially changing any of the
variables, disk mass, or gas disk lifetime, so that embryo growth
happens under similar gas conditions throughout the disk would promote
this outcome. However, for the Nominal conditions here - all typical
in the literature - the gas disk timescale is similar to embryo growth
near 1~au, leaving the outer regions of the disk to grow in
subtantially less gas-rich condtions.

The small Mars problem is otherwise dealt with in the literature by
significant involvment of the giant planets
\citep{Raymond:2009p6769,Walsh:2011p12463,Clement:2018}, non-smooth
distributions of solid materials \citep{Izidoro:2014} or vastly
different modes of accretion \citep{Levison:2015p20212}. Some of these
models rely on scenarios and simulations built on the same assumptions
assailed throughout this work - a simplistic bi-modal distribution of
embryos and planetesimals. For example, the ``Grand Tack'' scenario
modeled the migration of Jupiter through this simplistic disk, whereas
the results here suggest that there could be significant differences
in outcomes as a function of time and growth of the disk beyond
$\sim$~2~au (see \citealt{Jacobson:2014p18340,Brasser:2016}) - such that
the migrating giant planets encounter only planetesimals and no
embryos, or the opposite, during their migration.

Other avenues to improve and expand on this work lie in the
collisional model utilized. Here, using the \citet{Benz:1999p505}
disruption model was realtively simple, but there are more recent
investigations into the complex array of possible outcomes during
small and large impacts
\citep{Leinhardt:2012,Leinhardt:2015p18851,Movshovitz:2016p22086}. Given
the huge number of collisions between planetesimals means that small
shifts in disruptions laws at this size could add to big differences,
and the complexity of possible outcomes between embryos makes it hard
to generalize these critical final accretion/disruption
events. Improved modeling of collisions at both could change the
outcomes of models similar to this.

Other consequences that could spring from this work relate to
  rapid depletion of planetesimals near 1~au. The innermost
  planetesimals are accreted nearly to entirety and almost entirely
  into planets forming nearby. While there are some leftover
  planetesimals, there is minimal leftover mass relative to the
  typical bi-modal distributions typically used in modeling
  approaches. Whether the relative contribution to the planets from
  different regions of the inner solar system have substantially
  changed due to this affect is beyond the scope of the resolution of
  the simulations presented here and would require a specialized
  study.

  The other stark difference with some previous work is the
  quasi-stable phase of the first suite of planetary embryos, dubbed
  ``two generation of planets'' in the main text. The evolution
  described here would generate a different history of giant impacts
  between planetary embryos, with much of the actual ``giant impact''
  phase not kicking off until the nebular gas has greatly dissipated,
  potentially $\sim$~10~Myr later than previously modeled. Presuming
  that this is a generic effect, that even minimal amounts of nebular
  gas can provide stability for a system of planetary embryos, then it
  could change the way we look for or interpret signs of giant impacts
  in extra-solar planetary systems.

\section*{Acknowledgements}
KJW and HL were supported by NASA’s SSERVI program (Institute for the
Science of Exploration Targets) through institute grant number
NNA14AB03A, and NASA's Emerging Worlds program. This work used the
Extreme Science and Engineering Discovery Environment (XSEDE), which
is supported by National Science Foundation grant number ACI-1053575.

\bibliography{biblio}
\bibliographystyle{aasjournal}

\end{document}